\documentclass[twocolumn,american,english,amsmath,amssymb,prl,showpacs,superscriptaddress]{revtex4}
\usepackage[T1]{fontenc}
\usepackage[latin9]{inputenc}
\usepackage{color}
\usepackage{amsmath}
\usepackage{graphicx}
\usepackage{amssymb}
\usepackage{stmaryrd}
\usepackage{babel}

\def\kbar{{\mathchar'26\mkern-9mu k}}
\def\pred{\mathfrak{p}}

\begin{document}

\title{Experimental Test of Universality of the Anderson Transition}

\author{Matthias Lopez}

\affiliation{Laboratoire de Physique des Lasers, Atomes et Mol\'ecules, Universit\'e
Lille 1 Sciences et Technologies, CNRS; F-59655 Villeneuve d'Ascq
Cedex, France}

\homepage{www.phlam.univ-lille1.fr/atfr/cq}

\author{Jean-Fran\c cois Cl\'ement}

\affiliation{Laboratoire de Physique des Lasers, Atomes et Mol\'ecules, Universit\'e
Lille 1 Sciences et Technologies, CNRS; F-59655 Villeneuve d'Ascq
Cedex, France}

\author{Pascal Szriftgiser}

\affiliation{Laboratoire de Physique des Lasers, Atomes et Mol\'ecules, Universit\'e
Lille 1 Sciences et Technologies, CNRS; F-59655 Villeneuve d'Ascq
Cedex, France}

\author{Jean Claude Garreau}

\affiliation{Laboratoire de Physique des Lasers, Atomes et Mol\'ecules, Universit\'e
Lille 1 Sciences et Technologies, CNRS; F-59655 Villeneuve d'Ascq
Cedex, France}

\author{Dominique Delande}

\affiliation{Laboratoire Kastler-Brossel, UPMC-Paris 6, ENS, CNRS; 4 Place Jussieu,
F-75005 Paris, France}

\date{\today}
\begin{abstract}
We experimentally test the universality of the Anderson three dimensional metal-insulator transition. 
Nine sets of parameters controlling the microscopic details of this second order phase transition have been tested. 
The corresponding critical exponents are independent (within $2\sigma$) of these microscopic details,
 and the average value $1.63\pm0.05$ is in very good agreement with the numerically predicted value, $\nu = 1.58$.
\end{abstract}


\maketitle
In the presence of a disordered potential, the classical diffusive
transport of a particle can be inhibited by quantum interference among
the various paths where the particle is multiply scattered by disorder,
a puzzling phenomenon known as Anderson localization~\cite{Anderson:PR58}.
The dimensionality of the system plays a major role, which can be
understood qualitatively from the scaling theory of localization~\cite{Abrahams:PRL79}.
In dimension $d=3$ and above, there is a delocalized-localized (or
metal-insulator in solid-state physics language) transition --- known
as the Anderson transition. An energy ``mobility edge'' $E_{c}$,
which is a decreasing function of disorder, separates localized
motion at low energy from diffusive motion at high energy. On the
localized side, the localization length $\xi$ diverges algebraically
$\xi(E)\propto(E_{c}-E)^{-\nu},$ with $\nu$ the critical exponent
of the transition. On the diffusive side, the diffusion constant vanishes
like $D(E)\propto(E-E_{c})^{s}$ with, according to the scaling theory,
$s=(d-2)\nu$ \cite{Wegner:ZFP76,Abrahams:PRL79}. A key prediction
of the scaling theory is that the critical exponents are \textbf{universal},
that is, they do not depend on the microscopic details of the system,
such as the correlation functions of the disorder, the dispersion
relation of the particles, etc. Numerical experiments on simple models~\cite{Slevin:PRL99,Schreiber:EPJ00,Lemarie:EPL09}
such as the tight-binding Anderson model, have confirmed this universality,
with a non-trivial value of the critical exponent around $\nu=1.58$
for spinless time-reversal invariant 3-dimensional (3D) systems~\cite{Mirlin:RevModPhys08}.
However, there is a huge lack of experimental results.
In this letter, we present accurate measurements of the critical exponent
$\nu$ in 3D, and, by varying the various experimental parameters,
test whether its value is universal.

Anderson localization is due to interference between long multiple
scattering paths and is thus very sensitive to any mechanism destroying
the phase coherence of the wavefunction, making its experimental observation
and characterization very difficult~\cite{Janssen:PhysRep98}. In
the context of electronic transport in disordered samples, electron-electron
interaction is sufficiently important to partly invalidate the one-body
Anderson scenario, leading to a critical exponent close to unity~\cite{Katsumoto:1987}.
In a slightly different context, Anderson localization of acoustic
\cite{Page:NatPhys08} and electromagnetic \cite{Genack:N00,Wiersma:N97,Maret:PRL06}
waves has been experimentally observed. There, absorption is a limitation,
and the critical exponent $\nu\approx0.5$ estimated in \cite{Maret:PRL06}
does not agree with the numerical simulation result 1.58.

Cold atomic gases are conceptually simple systems which can be exposed
to well controlled disorder. For sufficiently cold samples, the phase
coherence time of the wavefunction describing the center of mass motion
can be kept sufficiently long for atomic interferometry experiments
to be routinely performed. Atom-atom interaction can be reduced to
a minimum, either by using dilute cold gases, or by using a Feshbach
resonance~\cite{Chin:Feshbach:RMP10}. Direct observation of Anderson
localization of atomic matter waves \emph{in a one-dimensional} disordered~\cite{Bouyer:N08}
or quasi-periodic~\cite{Inguscio:N08} potential has been reported,
the disordered potential being created by the effective interaction
with a detuned laser beam with a random spatial profile (speckle).
Very recently, observations of the Anderson localization in a 3D atomic gases have 
been claimed~\cite{ColdGases3D}. However, the interpretation of both results involves
heuristic assumptions, and a better understanding is needed before the observation of 
a mobility edge can be asserted.

This difficulty prompted the use of a slightly different system, where
the disordered potential is replaced by classical chaotic dynamics.
In the kicked rotor, a paradigmatic system of quantum chaos, quantum
mechanical interference tend to suppress the classical chaotic diffusive
dynamics, and to induce a phenomenon originally called \emph{dynamical
localization}, later discovered to be an analog of 1D Anderson localization
\emph{in momentum space}, by mapping the kicked rotor onto a quasirandom
1D Anderson model~\cite{Fishman:Altland}.
The experimental realization of the kicked rotor with laser-cooled
atoms interacting with a pulsed standing wave allowed the first experimental
observation of Anderson localization in 1D with atomic matter waves
\cite{Raizen:PRL94}. In order to observe the Anderson \emph{transition},
however, an analog of the \emph{3D} Anderson model is needed~\cite{GarciaGarcia:PRE09}. 
Here, we focus on the so-called ``quasi-periodic kicked rotor''
in which three incommensurate frequencies are used to generate the
3D character~\cite{Shepelyansky:PRL89}, and shown in~\cite{Lemarie:PRA09}
to be strictly equivalent to an anisotropic 3D Anderson model, a fact
further confirmed by a low-energy effective field theory~\cite{Tian:11}.
Meticulous numerical experiments~\cite{Lemarie:EPL09} have shown
the universality of the critical exponent for this system, $\nu=1.58\pm0.02,$
in excellent agreement with values found in the literature for the
Anderson model~\cite{Slevin:PRL99}.

An experiment based on this system has characterized the Anderson
metal-insulator transition~\cite{Chabe:PRL08}, with the first experimental
determination of the critical exponent $\nu=1.4\pm0.3.$ The experimental
setup, described in detail in~\cite{Lemarie:PRA09} consists of laser-cooled cesium
atoms interacting with a pulsed far-detuned standing wave (wavenumber
$k_{L}=7.4\times10^{6}$ m$^{-1}$ and maximum one-way intensity 180
mW). The amplitude of the kicks is modulated with two frequencies
$\omega_{2}$ and $\omega_{3}$. The Hamiltonian reads: 
\begin{equation}
H=\frac{p^{2}}{2}+K\cos x\left[1+\varepsilon\cos\left(\omega_{2}t\right)\cos\left(\omega_{3}t\right)\right]\sum_{n=0}^{N-1}\delta(t-n)\;,\label{eq:H}
\end{equation}
 where time is measured in units of the kicking period $T_{1}$, space
in units of $(2k_{L})^{-1}$, momentum in units of $2\hbar k_{L}/\kbar$,
with $\kbar=4\hbar k_{L}^{2}T_{1}/M$ ($M$ is the atom mass) playing
the role of an effective Planck constant ($[x,p]=i\kbar$) and $K$
is the average kick amplitude. The kicks are short enough (duration
$\tau=0.8$ $\mu$s) compared to the atom dynamics to be considered
as Dirac delta functions. If $\omega_{2}$, $\omega_{3},$ $\pi$
and $\kbar$ are incommensurate, this 1D quasiperiodic kicked rotor
has a 3D Anderson metal-insulator transition, displaying
localization in \emph{momentum space}.
Compared to~\cite{Chabe:PRL08} and~\cite{Lemarie:PRA09}, the
signal to noise and the stability of the experiment have been greatly
improved. Atomic momentum is measured by Raman stimulated
transitions. The previous Raman frequency generation setup
used direct current modulation of a master laser diode to drive the
Raman slave lasers~\cite{AP:DiodeMod:EPJD99}. This system has been
replaced by a fibered phase modulator driven at 9.2 GHz. Moreover,
3 (of 4) master diode lasers working with an external cavity setup
have been replaced by distributed feedback lasers, extending the experiment
stability from a few hours to several days. These improvements lead to much better experimental
signals, making
the determination of the critical exponent much more accurate and reliable, opening
the possibility to test its universality.

\begin{table}[t]
\begin{centering}
\begin{tabular}{c|c|c|c|c|c|c}
 & $\kbar$  & $\frac{\omega_{2}}{2\pi}$  & $\frac{\omega_{3}}{2\pi}$  & Path in ($K,\varepsilon$)  & $K_{c}$  & $\mathbf{\nu}$\tabularnewline
\hline
\hline
A & 2.89  & $\sqrt{5}$  & $\sqrt{13}$  & 4,0.1 $\shortrightarrow$ 8,0.8  & 6.67  & 1.63$\pm$0.06 \tabularnewline
\hline
B & 2.89  & $\sqrt{7}$  & $\sqrt{17}$  & 4,0.1 $\shortrightarrow$ 8,0.8  & 6.68  & 1.57$\pm$0.08 \tabularnewline
\hline
C & 2.89  & $\sqrt{5}$  & $\sqrt{13}$  & 3,0.435 $\shortrightarrow$ 10,0.435  & 5.91  & 1.55$\pm$0.25 \tabularnewline
\hline
D & 2.89  & $\sqrt{5}$  & $\sqrt{13}$  & 7.5,0 $\shortrightarrow$ 7.5,0.73  & $\varepsilon_{c}$=0.448  & 1.67$\pm$0.18\tabularnewline
\hline
E & 2.00  & $\sqrt{5}$  & $\sqrt{13}$  & 3,0.1 $\shortrightarrow$ 5.7,0.73  & 4.69  & 1.64$\pm$0.08 \tabularnewline
\hline
F & 2.31  & $\sqrt{5}$  & $\sqrt{13}$  & 4,0.1 $\shortrightarrow$ 9,0.8  & 6.07  & 1.68$\pm$0.06 \tabularnewline
\hline
G & 2.47  & $\sqrt{5}$  & $\sqrt{13}$  & 4,0.1 $\shortrightarrow$ 9,0.8  & 5.61  & 1.55$\pm$0.10 \tabularnewline
\hline
H & 3.46  & $\sqrt{5}$  & $\sqrt{13}$  & 4,0.1 $\shortrightarrow$ 9,0.8  & 6.86  & 1.66$\pm$0.12 \tabularnewline
\hline
I & 3.46  & $\sqrt{5}$  & $\sqrt{13}$  & 4,0.1 $\shortrightarrow$ 9,0.8  & 7.06  & 1.70$\pm$0.12 \tabularnewline
\end{tabular}
\par\end{centering}

\caption{\label{table:set-of-parameters} The 9 sets of parameters used: $\kbar$,
$\omega_{2}$ and $\omega_{3}$ control the microscopic details of
the disorder, $K$ controls the amplitude and $\epsilon$
the anisotropy of the hopping amplitudes. The critical point $K_{c}$
depends on the various parameters but the critical exponent is universal.
The weighted mean of the critical exponent is $\nu=1.63\pm0.05$.
The duration of the kicks is $\tau=0.8\,\mu$s for
sets A-H, and $\tau=0.96\,\mu$s for set I. }
\end{table}

A Sisyphus-boosted magneto-optical trap prepares
an initial thermal state of FWHM $4\times2\hbar k_{L}$,
much narrower than the final (localized or diffusive) momentum
distribution, and -- by time-reversal symmetry -- $\langle p(t)\rangle$
remains zero at all time $t$. We can directly monitor the dynamics
rather than rely on {}``bulk'' quantities such as the conductance,
itself related to the diffusion constant. A good quantity characterizing
the dynamics is $\langle p^{2}(t)\rangle.$ For practical and
historical reasons, the atomic momentum $P$ is measured in units
of two recoil momenta:
\begin{equation}
\pred=\frac{p}{\kbar}=\frac{P}{2\hbar k_{L}}.
\end{equation}
In the diffusive regime, $\langle\pred^{2}(t)\rangle$ increases linearly
with time; in the localized regime, it saturates at long times to
a value proportional to the square of the localization length. In~\cite{Lemarie:PRA09},
it was shown that, in order to characterize the Anderson transition,
one should measure: 
\begin{equation}
\Lambda(t)=\langle\pred^{2}(t)\rangle t^{-2/3}.\label{eq:Lambda}
\end{equation}
 $\Lambda(t)$ scales like $t^{-2/3}$ in the localized regime, like
$t^{1/3}$ in the diffusive regime and is \emph{constant} in the critical
intermediate regime. By measuring $\Lambda(t),$ one has thus a direct
access to the state of the system. $K$ is a control parameter playing
the role of the energy in the Anderson model. The critical value $K_{c}$
is a mobility edge separating a localized regime at low $K$ from
a diffusive regime at large $K$. The critical exponent $\nu$ can
be determined from the algebraic divergence $1/(K_{c}-K)^{\nu}$ of
the localization length near the critical point. For fixed $\kbar,\omega_{2},\omega_{3},$
one can also chose a path in the $(K,\varepsilon)$ plane which crosses
the critical line ``at right angle'', i.e.
faster, which improves the accuracy on the critical
exponent. Rather than measuring the full momentum distribution to
obtain $\langle\pred^{2}(t)\rangle,$ we measure the atomic population
in the zero-momentum class $\Pi_{0}(t).$ As shown in~\cite{Lemarie:PRL10},
$\langle\pred^{2}(t)\rangle$ is, by conservation of the number of
atoms, proportional to $\Pi_{0}^{-2}(t).$ The precise proportionality
factor depends on the shape of the momentum distribution, which has
been shown experimentally~\cite{Lemarie:PRL10} to vary smoothly
across the transition. The quantity measured experimentally, $\Lambda_{\mathrm{exp}}(K,t)=\Pi_{0}^{-2}(K,t)t^{-2/3}$,
has the same critical point and the same critical exponent than Eq.~(\ref{eq:Lambda}).
In the simplest experiment, parameters $\kbar,\omega_{2},\omega_{3},\varepsilon$
are fixed and $\Pi_{0}(K,t)$ recorded for several values of the
control parameter $K$, from which $\Lambda_{\mathrm{exp}}(K,t)$
is obtained.

\begin{figure}
\centering{}\includegraphics[width=0.9\columnwidth]{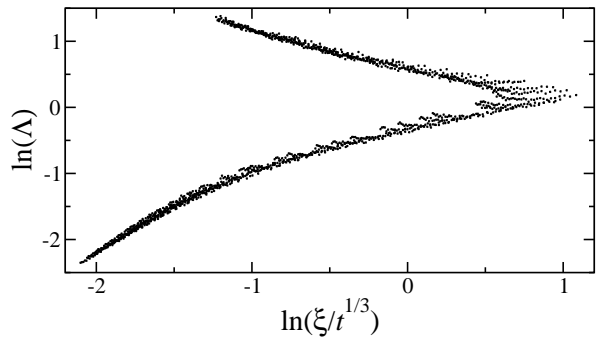} 
\caption{\label{fig:scaling} Typical scaling function constructed from the
experimental data measured on the quasi-periodic kicked rotor (set
E of Tab.~\ref{table:set-of-parameters}). The tip on the right
side indicates the critical point of the Anderson transition, the diffusive
(resp. localized) regime is the upper (resp. lower) branch. The dispersion
of the points is an indication of the uncertainty on the measured
data. }
\end{figure}

\begin{figure}
\centering{}\includegraphics[width=0.9\columnwidth]{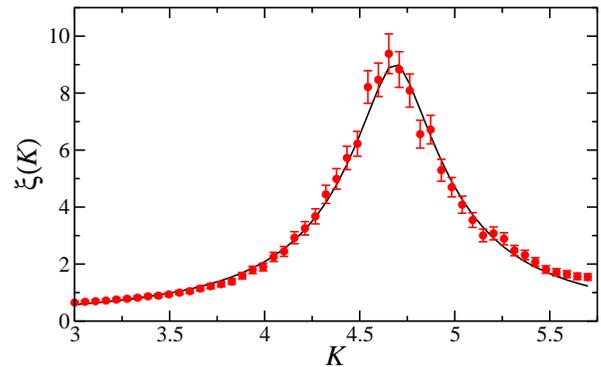} 
\caption{\label{fig:xsi} (color on line) Red points with error bars: Characteristic
length $\xi(K)$ constructed from the experimental data (set E of
Tab.~\ref{table:set-of-parameters}). At low $K$ (localized regime),
it is proportional to the localization length. At large $K$ (diffusive
regime), it is inversely proportional to the diffusion constant. The
``divergence" near $K_{c}=4.69$ is a signature of a second order quantum
phase transition. Several phenomena limit this divergence, see text
for discussion. They are taken into account by a fit including a cut-off
parameter, Eq.~(\ref{eq:fit}), shown as a solid curve, from which
the critical exponent $\nu$ is extracted.}
\end{figure}

As one cannot pursue the experiment for arbitrary long times, making
the distinction between the true critical point where $\Lambda_{\mathrm{exp}}(K_{c},t)$
is a constant and a neighboring point of small positive (localized)
or small negative (diffusive) slope is difficult. To circumvent
this, we use the finite-time scaling procedure explained in~\cite{Lemarie:PRA09},
which is a simple extension to the time domain of the finite-size
scaling method used in solid-state physics~\cite{MacKinnon:Pichard}.
We assume that there exists a scaling function $F$ such that: 
\begin{equation}
\Lambda_{\mathrm{exp}}(K,t)=F\left(\xi(K)t^{-1/3}\right)
\end{equation}
where $\xi(K)$ is an unknown function to be determined. The idea,
described in detail in~\cite{Lemarie:PRA09,Lemarie:These:09} is
that $\xi(K)$ should be chosen so that $F\left(\xi(K)t^{-1/3}\right)$
is an continuous function, safe for the divergence at the critical
point, with two distinct localized and diffusive branches. In order
to do so, we try to collapse the various experimental curves $\Lambda_{\mathrm{exp}}(K,t)$,
corresponding to each value of $K$, in a single curve, the various
$\xi(K)$ - associated to the localization length 
on the localized branch - being free fitting parameters. A typical
scaling function is shown in Fig.~\ref{fig:scaling} and the corresponding
fit parameter function $\xi$ in Fig.~\ref{fig:xsi}. Comparison
with our previously published data, e.g. Fig.~3(c) in~\cite{Chabe:PRL08},
shows a dramatic improvement in the quality of the measurements.
Each point $\Lambda_{\mathrm{exp}}(K,t)$ results from independent
experiments, making the statistical errors rather easy to evaluate.
We checked that the dispersion of points around the scaling function
in Fig.~\ref{fig:scaling} is of the order of $\Delta(\ln\Lambda)\approx0.05,$
limited by the experimental uncertainties, the $\chi^{2}$ per degree
of freedom of the fit being typically slightly smaller than 1.

From finite-time measurements, it is impossible
to extract a truly diverging $\xi(K)$. Several spurious phenomena~\cite{Lemarie:PRA09}
are responsible for the observed cut-off: The dominant one is the
finite duration of the experiment. Several sources of decoherence,
such as spontaneous emission, collisions between atoms, residual effect
of gravity because the laser beams are not perfectly horizontal, and
the inhomogeneity related to the Gaussian intensity profile of the
standing wave contribute to the cut-off. We have
increased the phase coherence time up to 200-300 kick periods, in
agreement with theoretical calculations~\cite{Lemarie:PRA09}. In
order to reduce uncontrolled systematic effects, we chose to use the
same time interval - from 30 to 150 kicks - for the analysis of all
experimental data. The finite duration is also responsible for systematic
effects: It tends to slightly underestimate $K_{c},$ but does not
seem to shift significantly the critical exponent. Using a different
interval, 30-120 kicks, produces $\nu$ values not differing by more
than 0.05.

A typical $\xi(K)$ curve, Fig.~\ref{fig:xsi}, displays a clear
``divergence'' near the critical point (increase by more than
one order of magnitude, much better than in~\cite{Chabe:PRL08}).
It is itself fitted to extract the position of the critical point
$K_{c}$ and the critical exponent $\nu,$ using the following formula:
\begin{equation}
1/\xi(K)=\alpha|K-K_{c}|^{\nu}+\beta\label{eq:fit}
\end{equation}
where $\beta$ is a cut-off parameter taking into account the various
limitations discussed above. As seen in Fig.~\ref{fig:xsi} the fit
is excellent. The fitting parameter $\nu$ depends on the range of
$K$ where the fit is performed. A too small range produces a large
uncertainty in $\nu$ while the quality of the fit deteriorates for
a too large range. To avoid any bias, we have fitted all data sets
in the interval $[0.8K_{c},1.2K_{c}]$, for which the $\chi^{2}$
per degree of freedom is of the order of 1.
The uncertainties are calculated using a bootstrap method
starting from the raw experimental data and their error bars, ending
to the determination of $\nu$ through the construction of the scaling
function. For data sets presented, no statistically significant anomaly
has been detected. The statistical uncertainty on $K_{c}$ is 
very small, at most few $10^{-2},$ but the finite duration of the
experiment is responsible for a systematic shift towards low $K$
(see above).

In order to test the universality of the critical exponent, we have
chosen a ``reference'' set of parameters, noted A in table~\ref{table:set-of-parameters},
which has the same parameters used in~\cite{Chabe:PRL08}.
We have then modified the $\omega_{2}$ and $\omega_{3}$ frequencies
for set B. We have also modified the path in the $(K,\varepsilon)$
plane, either by changing $K$ only (set C) or $\varepsilon$ only
(set D). In these two situations, the crossing of the critical regime
is slower, making the accuracy on $\nu$ significantly worse. We have
further modified the kicking period $T_{1},$ which affects the effective
Planck's constant $\kbar$. Several smaller $\kbar$ values have been
used in sets E, F and G. One larger value of $\kbar$ was used in
sets H and I the difference between the two sets being the duration
of the laser pulses. The
fact that very close values of $K_{c}$ and the same critical exponent
are obtained is a strong indication that our experimental system is
well described by the model Hamiltonian, Eq.~(\ref{eq:H}).

\begin{figure}
\centering{}\includegraphics[width=0.9\columnwidth]{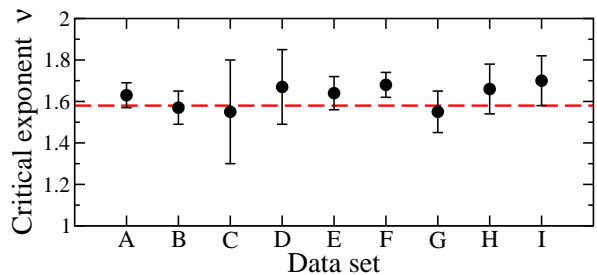}
 \caption{\label{fig:nu} (color online) Experimental test of the universality
of the metal-insulator transition. The critical exponent $\nu$, measured
for 9 different sets of parameters A-I, is \emph{universal}, i.e.
independent of the microscopic details. The error bars indicate one
standard deviation, measured using the experimental uncertainties
and a bootstrap technique. The horizontal dashed line is the commonly accepted value
$\nu=1.58,$ determined
from extensive numerical calculations on various models. Parameter
values: See Tab.~\ref{table:set-of-parameters}.}
\end{figure}

The final results are given in table~\ref{table:set-of-parameters}
and plotted with their error bars (one standard deviation) in Fig.~\ref{fig:nu}.
They unambiguously demonstrate the universality of the localized-diffusive
transition in the quasi-periodic kicked rotor. Moreover, all numerical
values are compatible (within 2 standard deviations) with the best
numerical determinations of $\nu=1.58\pm0.02,$ both for the kicked
rotor and the Anderson model. They all markedly differ from the value
1 predicted by the self-consistent theory of localization~\cite{Vollhardt:book91}.
The later theory is an attempt to justify the scaling properties using
a microscopic approach: It is qualitatively correct and gives simple
physical pictures. For example, it has been used to successfully predict the momentum
distribution at the critical point~\cite{Lemarie:PRL10}. However,
it lacks a key ingredient: At criticality, the wavefunctions display
very large fluctuations which can be characterized by a multifractal
spectrum~\cite{Mirlin:RevModPhys08,Martin:PRE10,Faez:PRL09}. Huge
fluctuations are known to affect critical exponents of thermodynamic
phase transitions, it is thus no surprise that they also affect the
Anderson transition. While quantum phase transitions are usually considered
for the ground state of the system, it must be emphasized that the
Anderson transition deals with excited states in the vicinity of the
mobility edge, displaying much richer properties. Especially, ultra-cold
atomic gases open the way to experimental studies of the interplay
of disorder, interference and interactions.
\begin{acknowledgments}
We thank G. Lemari\'e for useful discussions, and R. Holliger for help with the experiments.
\end{acknowledgments}
%

%\bibliographystyle{apsrev}
%\bibliography{univ1,D:/JCG/Publications/ArtDataBase}

\end{document}